\documentclass[aps,pra,twocolumn,showpacs,groupedaddress]{revtex4}

\usepackage{graphicx}
\newcommand{\Tr}{\text{Tr}}
\newcommand{\ket}[1]{|#1\rangle}
\newcommand{\bra}[1]{\langle#1|}

\newcommand{\schrodinger}{Schr\"{o}dinger}
\newcommand{\sz}{\sigma^z}

\newcommand{\sigmam}{\sigma^-}
\newcommand{\sigmap}{\sigma^+}
\newcommand{\ext}[2]{|#1\rangle\langle#2|}   

\newcommand{\p}[1]{\frac{\partial}{\partial t}#1} 

\newcommand{\PRA}[3] {Phys. Rev. A {\bf #1}, #2
(#3)}
\newcommand{\PRB}[3] {Phys. Rev. B {\bf #1}, #2
(#3)}

\newcommand{\PRL}[3] {Phys. Rev. Lett. {\bf #1}, #2
(#3)}
\newcommand{\JPA}[3] {J. Phys. A {\bf #1}, #2 (#3)}
\newcommand{\JPB}[3] {J. Phys. B {\bf #1}, #2 (#3)}
\newcommand{\PLA}[3] {Phys. Lett. A {\bf #1}, #2 (#3)}
\newcommand{\EPL}[3] {EPL {\bf #1}, #2 (#3)}

\begin{document}
\title{Nonequilibrium thermal entanglement in three-qubit $XX$  model}
\author{X. L. Huang\footnote{ghost820521@163.com or huangxiaoli1982@gmail.com}}
\author{J. L. Guo}
\author{X. X. Yi\footnote{yixx@dlut.edu.cn}}
\affiliation{School of Physics and Optoelectronic Technology,\\
Dalian University of Technology, Dalian 116024 China }

\date{\today}

\begin{abstract}
Making use of the master equation and effective Hamiltonian
approach, we investigate the steady state entanglement in a
three-qubit $XX$ model. Both  symmetric and nonsymmetric qubit-qubit
couplings are considered.  The system (the three qubits) is coupled
to two bosonic baths at different temperatures. We calculate the
steady state by the effective Hamiltonian approach and discuss  the
dependence of the steady state entanglement on the temperatures and
couplings. The results show that for symmetric qubit-qubit
couplings, the entanglements between the nearest neighbor are equal,
independent of the temperatures of the two baths. The maximum
 of the entanglement  arrives  at $T_L=T_R$.
 For nonsymmetric qubit-qubit couplings, however, the situation is totally
 different. The baths at different temperatures
 would benefit the entanglement and the entanglements between the nearest neighbors
are no longer equal. By examining the probability distribution of
each eigenstate in the steady state, we present an explanation for
these observations. These results suggest that the steady
entanglement can be controlled by the temperature of the two baths.

\end{abstract}

 \maketitle

When we talk about a real physical system, we should take the
effects of its environment into account because all quantum systems
interact unavoidably with their surroundings. A quantum  system that
can not isolate from its environment is usually referred to  open
quantum system \cite{BPbookprojection}. The dynamics of open quantum
systems can be described by  quantum master equations in the
Schr\"odinger picture, or Langevin equation in the Heisenberg
picture \cite{quantumoptics}. The coupling of  environment to a
system definitely changes the properties of the system, such as the
geometric phase \cite{Berry} and entanglement
\cite{Entanglement1935}.

Entanglement is a quantum  resource that  has no classical
counterpart. It was first recognized in 1935, and  has been widely
studied in recent years due to its potential applications in quantum
information processing \cite{nielsen}. The environment can  either
induce entanglement \cite{induceE} or decrease entanglement (in this
case environment often lead to a death of  entanglement, which is
usually called entanglement sudden death \cite{ESD}). When a quantum
system is in contact with a heat reservoir at a fixed temperature,
the system will relax   into a thermal equilibrium state
$\rho(T){=}e^{-\beta H}/\Tr(e^{-\beta H})$ eventually. The
entanglement of this state is called thermal entanglement,
 which has been extensively studied  in the past decade \cite{thermalE}.

In this paper, we shall study a different type of thermal
entanglement. It depends on temperature but its state is not
statistically  equilibrium. This situation arises when a quantum
system interacts with two bosonic baths at different  temperatures.
The state of the quantum system eventually arrived is not a
statistically equilibrium but a steady state \cite{steadystateNon}.
In Ref.\cite{Yan0809PRB}, the heat transport were studied for such a
system. In Ref.\cite{NTE2007}, the entanglement
 of a two qubit $XX$ chain were studied for both identical and different qubits.
 In this paper, we will study the
entanglement  of a three qubit $XX$ chain coupled to two baths at
different temperatures. Two cases, i.e.,  symmetric and nonsymmetric
qubit-qubit couplings, are considered.

\begin{figure}
\includegraphics*[width=0.9\columnwidth]{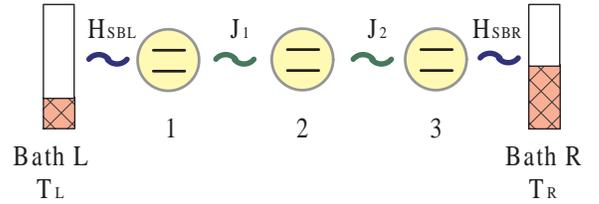}\caption{A schematic representation of our model.
 The two-level systems are connected to two bosonic  baths
 held at different temperatures, $T_L$ and $T_R$, respectively.}\label{FIG:setup}
\end{figure}

Consider a spin chain consisting of three spins with $XX$
interaction. The Hamiltonian for this system has the form,
\begin{eqnarray}
H{=}\sum_{n=1}^3\frac{\varepsilon}2\sz_n{+}J_1(\sigmap_1\sigmam_2{+}\sigmam_1\sigmap_2){+}J_2(\sigmap_2\sigmam_3{+}\sigmam_2\sigmap_3),
\end{eqnarray}
where $\sz_n$ and $\sigma^{\pm}_n$ are Pauli operators for the $n$th
spin, and $J_1$ and $J_2$ denote coupling constants. Spin 1 and 3
interact with two sperate bosonic baths at different temperatures
$T_L$ and $T_R$, respectively (see Fig.\ref{FIG:setup}). The
Hamiltonian for each bath $j{=}L,R$ is given by
$H_{Bj}{=}\sum_n\omega_{n,j}b^{\dag}_{n,j}b_{n,j}$ and the
interaction between the spin and its bath is described by
$H_{SBj}{=}\sigmap_j\sum_ng_n^{(j)}b_{n,j}{+}\sigmam_j\sum_ng_n^{(j)}b_{n,j}^{\dag}=
\sum_{\mu}X_{j,\mu}^+\Gamma_{j,\mu}{+}X_{j,\mu}^-\Gamma_{j,\mu}^{\dag}$
($j=1,3$ for spin), here the operator $X_{j,\mu}$ is an
eigenoperator of the system Hamiltonian satisfying
$[H,X_{j,\mu}^{\pm}]{=}\pm\omega_{j,\mu}X_{j,\mu}^{\pm}$, and
$\Gamma_{j,\mu}$ acts on the bath degrees of freedom.  We assume the
both baths are in the uncorrelated thermal equilibrium states. Then
the density operator of the bath is
$\rho_{Bj}{=}e^{-\beta_jH_{Bj}}/\Tr(e^{-\beta_jH_{Bj}})$.
In this paper we set $k_B{=}\hbar{=}1$ for simplicity. Under this
assumption, the dimension of the transition frequency $\varepsilon$,
coupling constants $J_1,J_2$ and the temperatures $T_L,T_R$ are
same. In our following discussion, the value of these parameters
only stand for the ration relation between them. The dynamics of the
system affected by the two baths can be described by the quantum
master equation \cite{quantumoptics}, which is obtained by tracing
out the bath variables within the Born-Markovian approximation. This
equation can usually be arranged in Lindblad form
\begin{eqnarray}
\p\rho=-i[H,\rho]+\mathcal{L}_L\rho+\mathcal{L}_R\rho,\label{eqn:masterequation}
\end{eqnarray}
where $\mathcal{L}_L\rho$ ($\mathcal{L}_R\rho$) is the dissipative
term due to the coupling of the system to the left (right) bath. In
order to find an exact form of the dissipative term, we go to the
bases composed by the  eigenstates $\ket{s_i}$ of the system
Hamiltonian
 with the corresponding eigenvalues $\lambda_i$
\begin{eqnarray}
  &&\ket{s_1}{=}\ket{111},\lambda_1{=}\frac32\varepsilon,\nonumber\\
  &&\ket{s_2}{=}\ket{000}, \lambda_2{=}-\frac32\varepsilon, \nonumber  \\
  &&\ket{s_3}{=}\frac{\sqrt{2}}2(\sin\theta\ket{110}{-}\ket{101}{+}\cos\theta\ket{011}), \lambda_3{=}\frac12\varepsilon{-}J,\nonumber \\
  &&\ket{s_4}{=}\cos\theta\ket{110}{-}\sin\theta\ket{011},\lambda_4{=}\frac12\varepsilon,\nonumber\\
  &&\ket{s_5}{=}\frac{\sqrt{2}}2(\sin\theta\ket{110}{+}\ket{101}{+}\cos\theta\ket{011}),\lambda_5{=}\frac12\varepsilon{+}J,\nonumber\\
  &&\ket{s_6}{=}\frac{\sqrt{2}}2(\cos\theta\ket{100}{-}\ket{010}{+}\sin\theta\ket{001}),\lambda_6{=}{-}\frac12\varepsilon{-}J,\nonumber\\
  &&\ket{s_7}{=}\sin\theta\ket{100}{-}\cos\theta\ket{001},\lambda_7{=}{-}\frac12\varepsilon,  \\
  &&\ket{s_8}{=}\frac{\sqrt{2}}2(\cos\theta\ket{100}{+}\ket{010}{+}\sin\theta\ket{001}),\lambda_8{=}{-}\frac12\varepsilon{+}J,\nonumber
\end{eqnarray}
where $J{=}\sqrt{J_1^2+J_2^2}$, and $\tan\theta{=}\frac{J_2}{J_1}$.
In this representation, the dissipative term $\mathcal{L}_j\rho$
($j{=}L,R$) can be written as
\begin{eqnarray}
\mathcal{L}_j\rho=\sum_{\mu=1}^3J^{(j)}(-\omega_{\mu})\left(2X_{j,\mu}\rho
X_{j,\mu}^{\dag}-\{\rho,X_{j,\mu}^{\dag}X_{j,\mu}\}\right)\nonumber\\
+J^{(j)}(\omega_{\mu})\left(2X_{j,\mu}^{\dag}\rho
X_{j,\mu}-\{\rho,X_{j,\mu}X_{j,\mu}^{\dag}
\}\right),\label{eqn:dissipativeterm}
\end{eqnarray}
with eigenfrequencies,
\begin{eqnarray}
\omega_1=\varepsilon+J,~~~~ \omega_2=\varepsilon,~~~~
\omega_3=\varepsilon-J,
\end{eqnarray}
and corresponding eigenoperators
\begin{eqnarray}
&&X_{L,1}=\frac{\sqrt{2}}2\cos\theta\ext{s_3}{s_1}-\frac12\ext{s_6}{s_4}\nonumber\\
&&~~~~~~~~~~~-\frac{\sqrt{2}}2\cos\theta\ext{s_7}{s_5}+\frac{\sqrt{2}}2\cos\theta\ext{s_2}{s_8},\nonumber\\
&&X_{L,2}=-\frac{\sqrt{2}}2\ext{s_4}{s_1}-\sin\theta\ext{s_6}{s_3}\nonumber\\
&&~~~~~~~~~~~+\sin\theta\ext{s_8}{s_5}+\sin\theta\ext{s_2}{s_7},\nonumber\\
&&X_{L,3}=\frac{\sqrt{2}}2\cos\theta\ext{s_5}{s_1}+\frac{\sqrt{2}}2\cos\theta\ext{s_7}{s_3}\nonumber\\
&&~~~~~~~~~~~+\frac12\ext{s_8}{s_4}+\frac{\sqrt{2}}2\cos\theta\ext{s_2}{s_6},\nonumber
\end{eqnarray}
\begin{eqnarray}
&&X_{R,1}=\frac{\sqrt{2}}2\sin\theta\ext{s_3}{s_1}+\frac12\ext{s_6}{s_4} \nonumber\\
&&~~~~~~~~~~~+\frac{\sqrt{2}}2\sin\theta\ext{s_7}{s_5}+\frac{\sqrt{2}}2\sin\theta\ext{s_2}{s_8},\nonumber\\
&&X_{R,2}=\frac{\sqrt{2}}2\ext{s_4}{s_1}-\cos\theta\ext{s_6}{s_3}\nonumber\nonumber\\
&&~~~~~~~~~~~+\cos\theta\ext{s_8}{s_5}-\cos\theta\ext{s_2}{s_7},\nonumber\\
&&X_{R,3}=\frac{\sqrt{2}}2\sin\theta\ext{s_5}{s_1}-\frac{\sqrt{2}}2\sin\theta\ext{s_7}{s_3}\nonumber\\
&&~~~~~~~~~~~-\frac12\ext{s_8}{s_4}+\frac{\sqrt{2}}2\sin\theta\ext{s_2}{s_6}.
\end{eqnarray}
The spectral density in Eq.(\ref{eqn:dissipativeterm}) is defined by
\begin{eqnarray}
J^{(j)}(\omega_{\mu})=\int_0^{\infty}d\tau
e^{i\omega_{j,\mu}}\left\langle
e^{-iH_{Bj}\tau}\Gamma^{\dag}_{j,\mu}e^{iH_{Bj}\tau}\Gamma_{j,\mu}\right\rangle.
\end{eqnarray}
In this paper, the bath is assumed as an infinite set of harmonic
oscillators, so the spectral density has the form
$J^{(j)}(\omega_{\mu}){=}\gamma_j(\omega_{\mu})n_j(\omega_{\mu})$,
where $n_j(\omega_{\mu}){=}(e^{\beta_j\omega_{\mu}}{-}1)^{-1}$ and
$J^{(j)}(-\omega_{\mu}){=}e^{\beta_j\omega_{\mu}}J^{(j)}(\omega_{\mu})$.
On the same footing as  the Born-Markovian approximation, we take
here a Weisskopf-Winger form for the coupling constant as
$\gamma_j(\omega_{\mu}){=}\gamma_j$, i.e., the coupling constant is
spectrum  independent.

The master equation can be solved by the fourth-order Runge-Kutta
method and the steady state can be reached when we set the evolution
time long enough \cite{Yan0809PRB}. Different from this method, we
solve the steady state of Eq.(\ref{eqn:masterequation}) here with
the help of effective Hamiltonian approach \cite{EHAall}. The main
idea of this method can be described as follows. By introducing an
ancilla, we map the density matrix and the master equation to a
state vector and a \schrodinger-like equation. The solution of the
master equation can be obtained by mapping the solution of the
\schrodinger-like equation back to the density matrix. Rewrite the
master equation in the Lindblad form \cite{Lindblad} as $\p
\rho{=}-i[H,\rho]{-}\frac12\sum_k\left(L_k^{\dag}L_k\rho{+}\rho
L_k^{\dag}L_k{-}2L_k\rho L_k^{\dag}\right)$. Then the effective
Hamiltonian is defined by
$\mathcal{H}_T=H{-}H^A{-}\frac{i}2\sum_kL_k^{\dag}L_k{-}\frac{i}2\sum_kL_k^{A\dag}L_k^A
{+}i\sum_kL_k^AL_k$, where $A$ denotes the operator for the ancilla,
which is defined by
$\bra{e_m}O^A\ket{e_n}{=}\bra{E_n}O^{\dag}\ket{E_m}$. Here
$\{\ket{E_n}\}$ and $\{\ket{e_n}\}$ are time independent bases for
the original system and the ancilla. The initial state independent
steady state corresponding to the eigenstate of $\mathcal{H}_T$ with
eigenvalue zero. Then we can study the steady state properties for
such a system.
\begin{figure}
\includegraphics*[width=0.45\columnwidth]{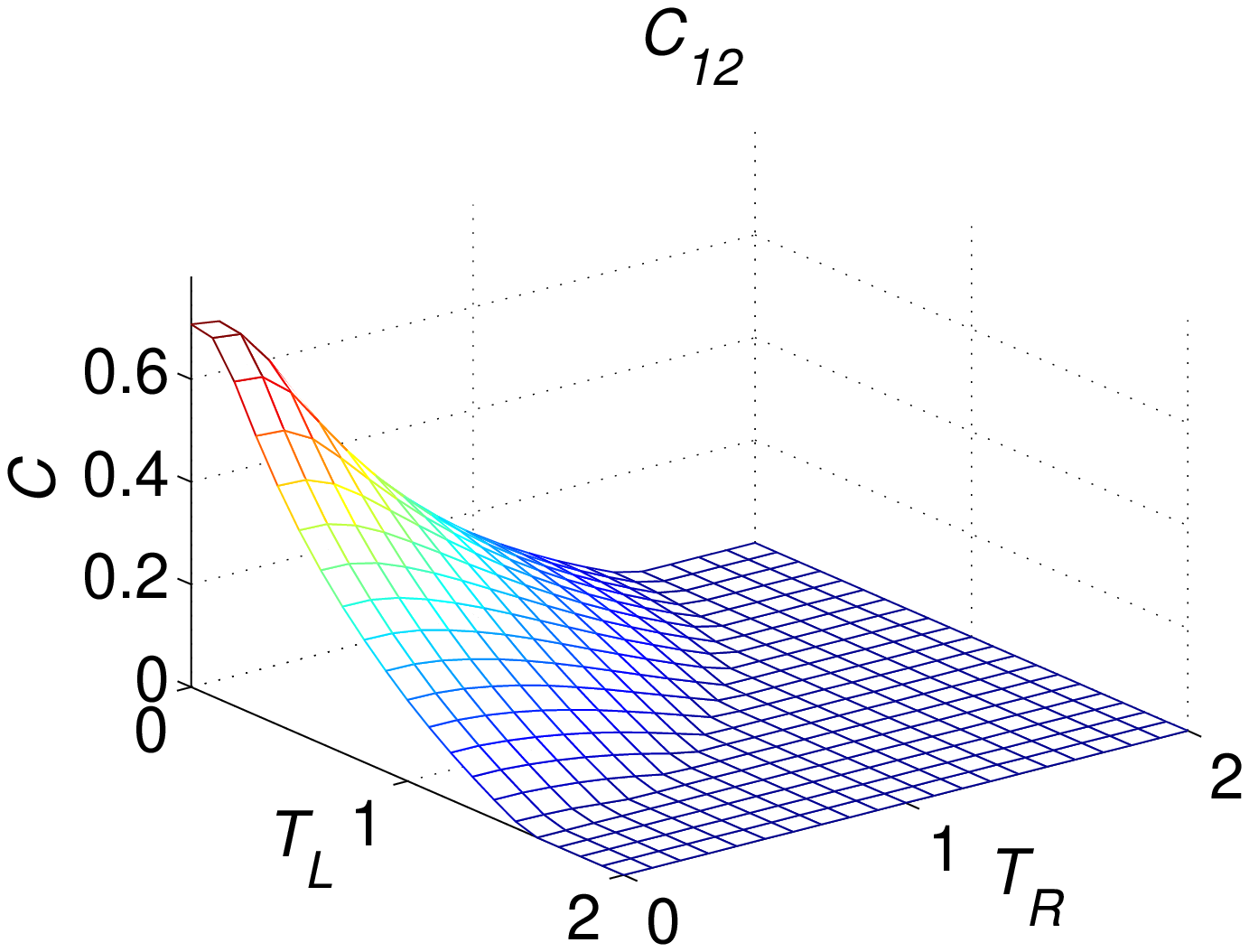}
\includegraphics*[width=0.45\columnwidth]{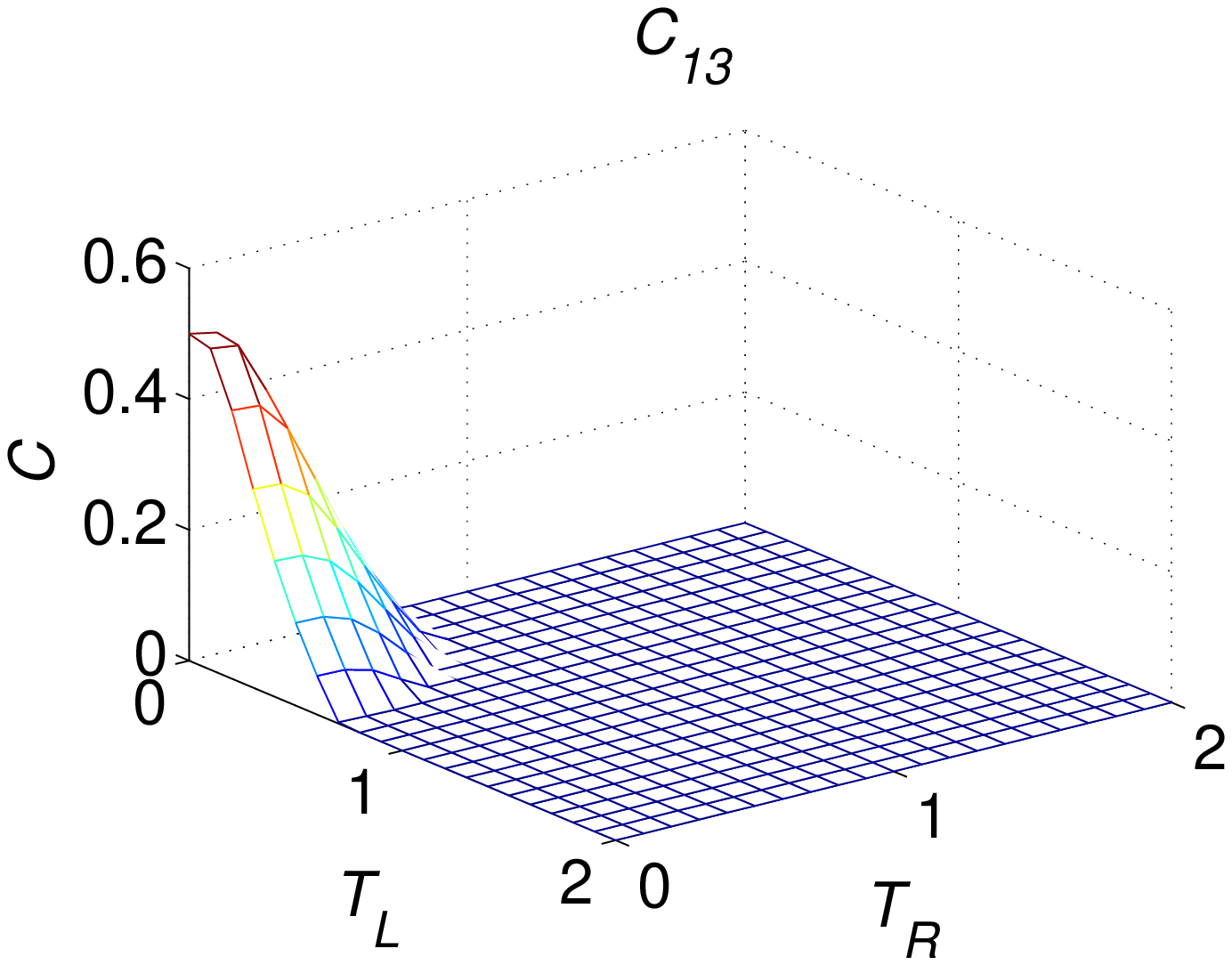}\caption{Steady state entanglement
$C_{12}$ and $C_{13}$ as functions of the bath temperatures $T_L$
and $T_R$. Here $\varepsilon{=}1$ and
$J_1{=}J_2=1$.}\label{FIG:Cw1J1}
\end{figure}

\begin{figure}
\includegraphics*[width=0.45\columnwidth]{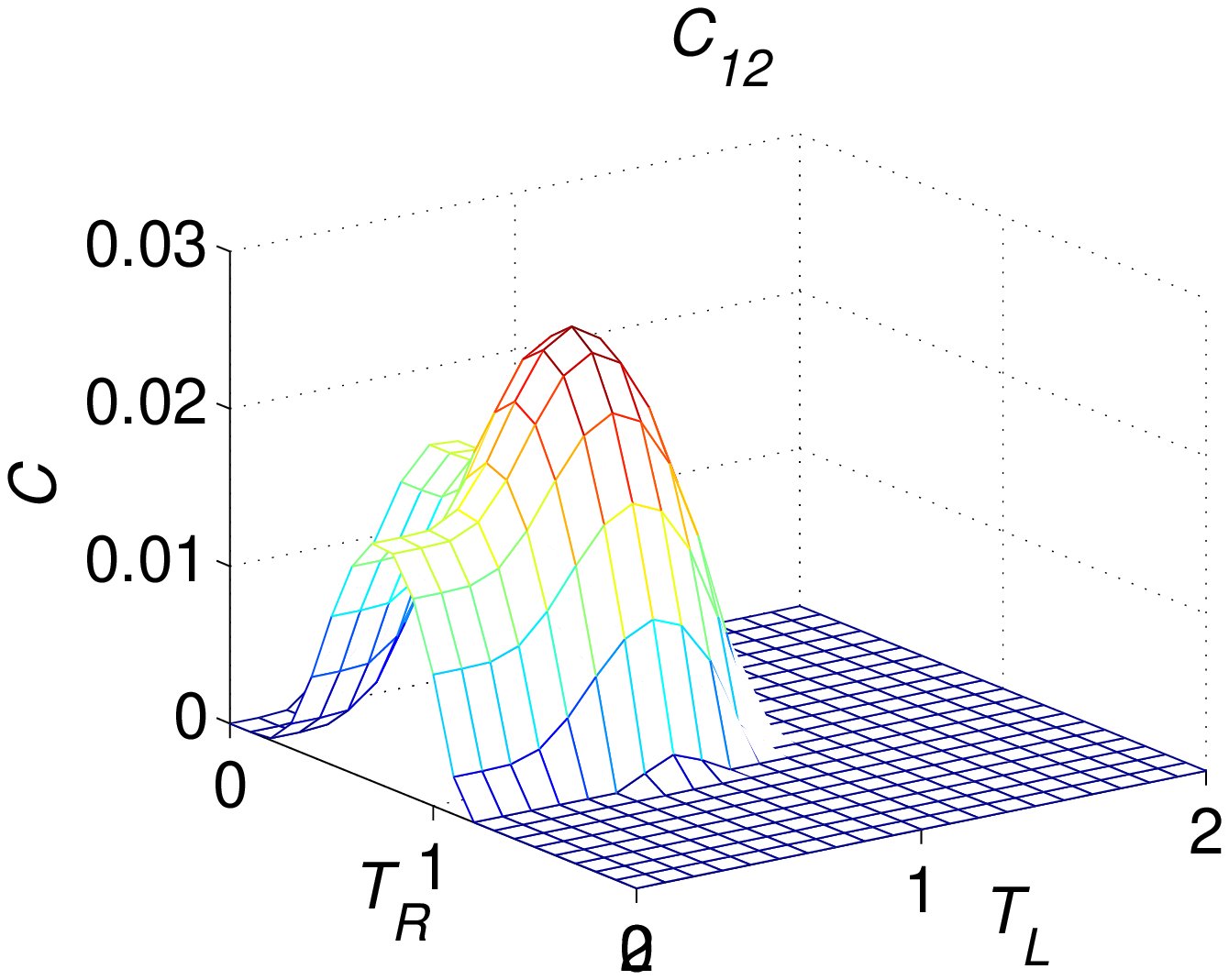}
\includegraphics*[width=0.45\columnwidth]{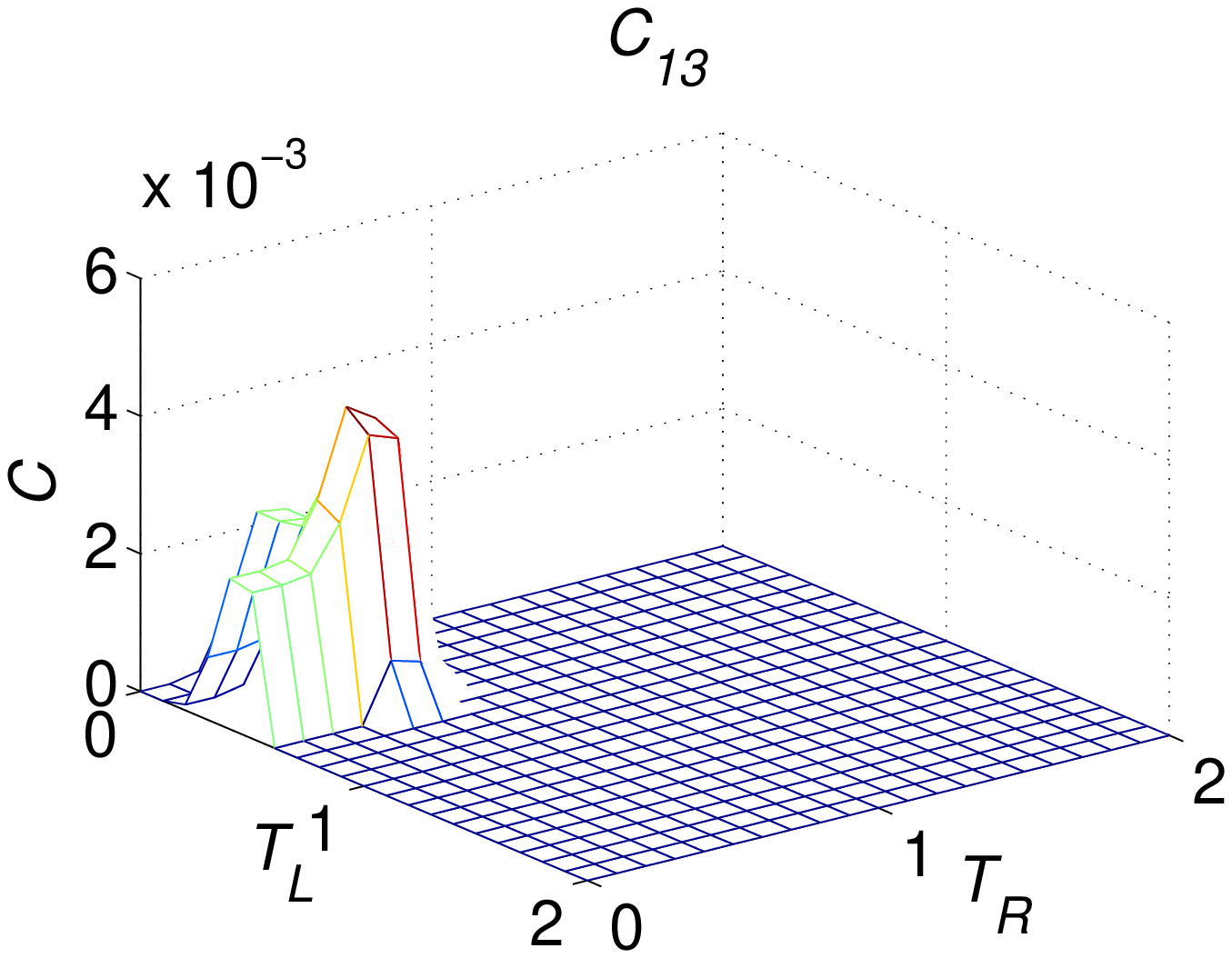}
\caption{Steady state entanglement $C_{12}$ and $C_{13}$ as
functions of the bath temperatures $T_L$ and $T_R$. Here
$\varepsilon{=}3$ and $J_1{=}J_2{=}1$.}\label{FIG:Cw3J1}
\end{figure}
\begin{figure}
\includegraphics*[width=0.45\columnwidth,height=0.25\columnwidth]{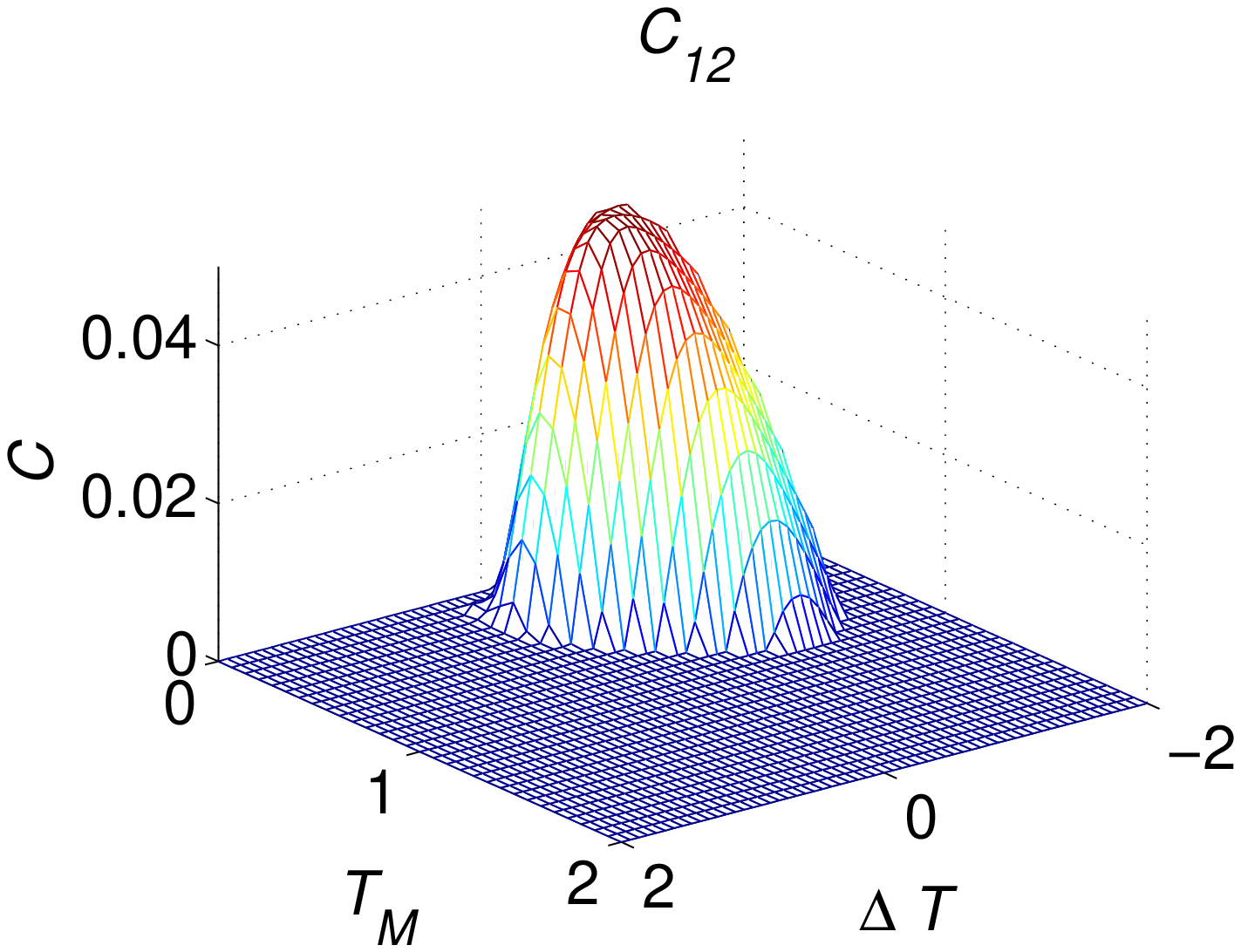}
\includegraphics*[width=0.45\columnwidth,height=0.25\columnwidth]{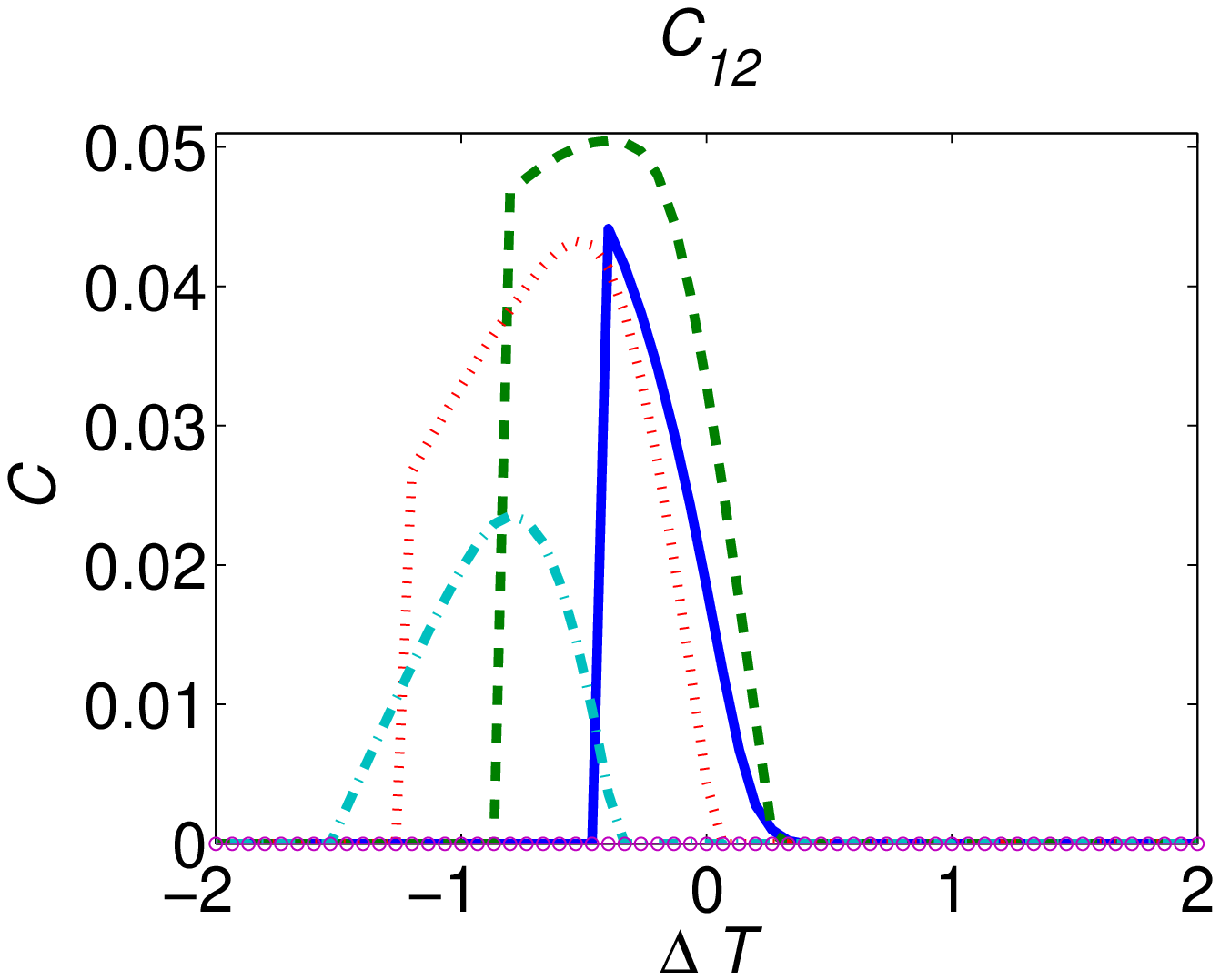}
\includegraphics*[width=0.45\columnwidth,height=0.25\columnwidth]{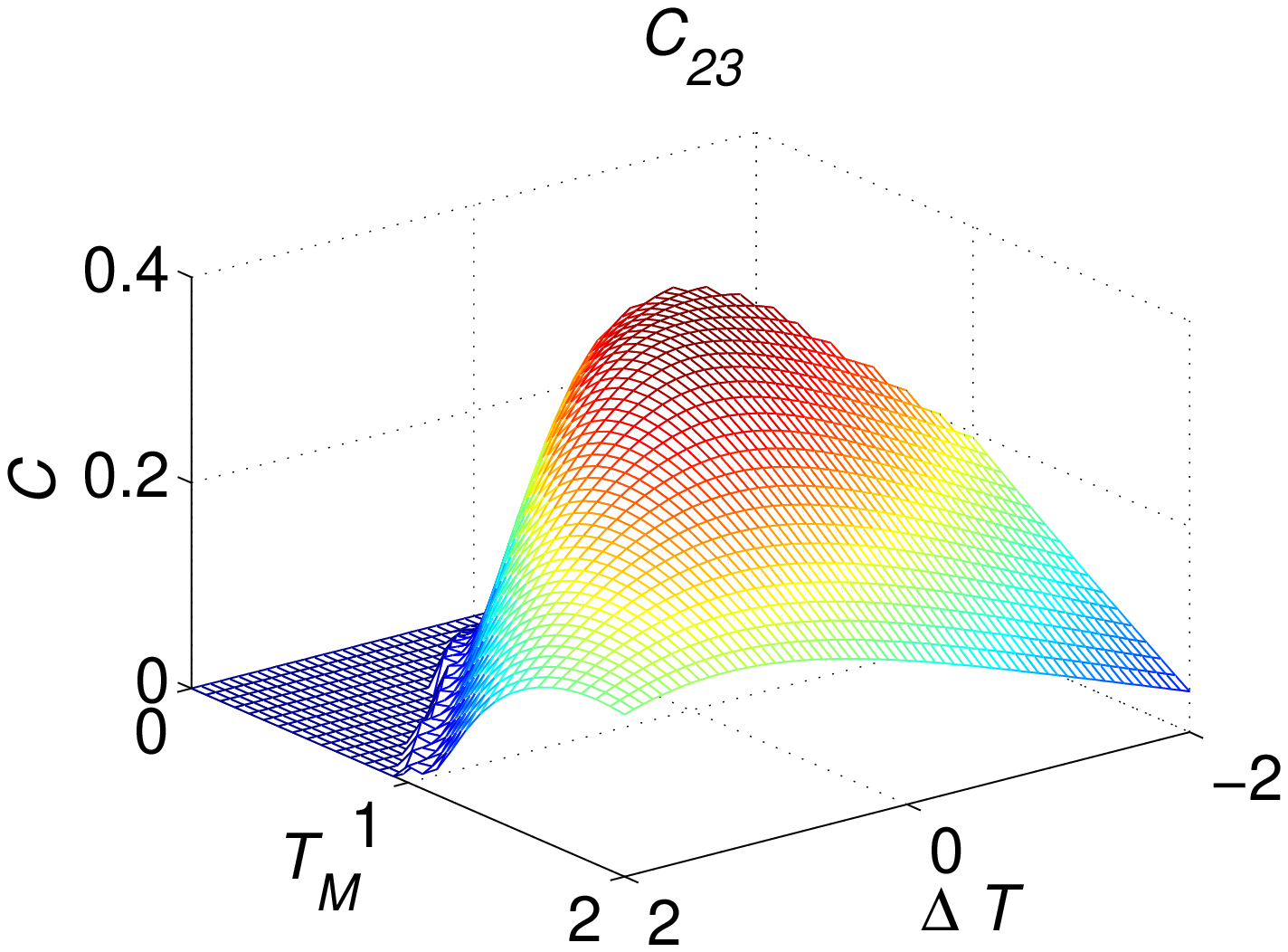}
\includegraphics*[width=0.45\columnwidth,height=0.25\columnwidth]{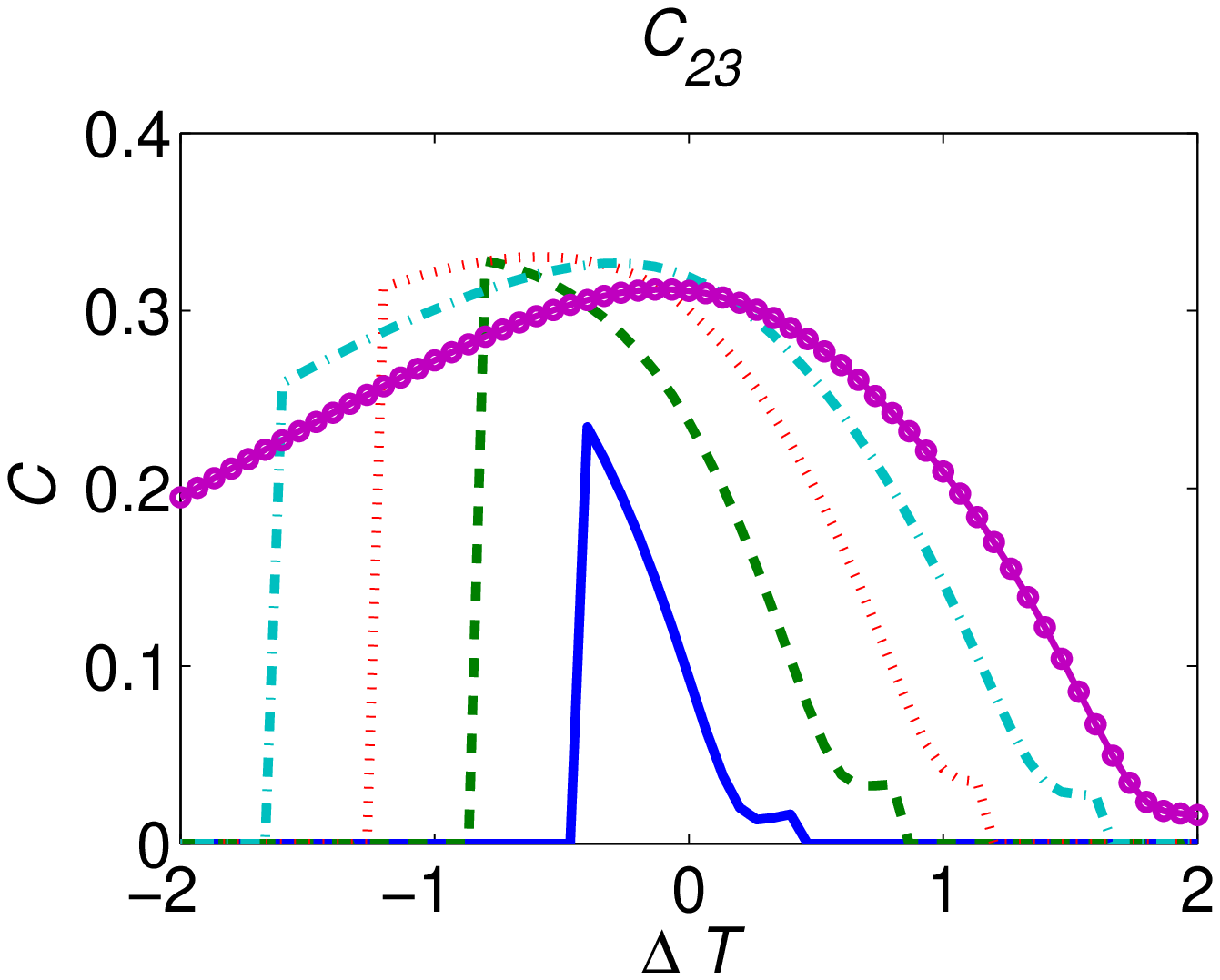}
\includegraphics*[width=0.45\columnwidth,height=0.25\columnwidth]{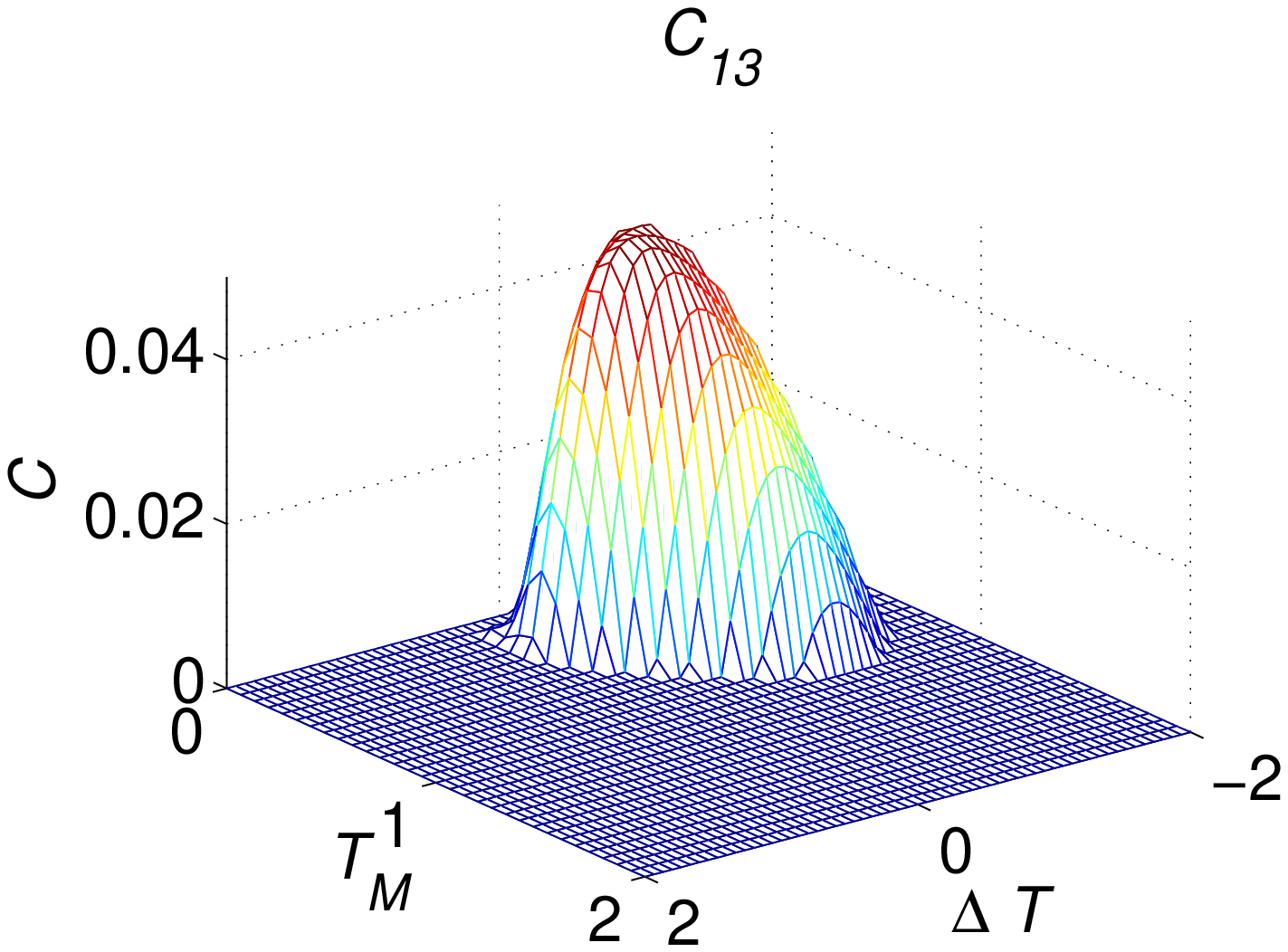}
\includegraphics*[width=0.45\columnwidth,height=0.25\columnwidth]{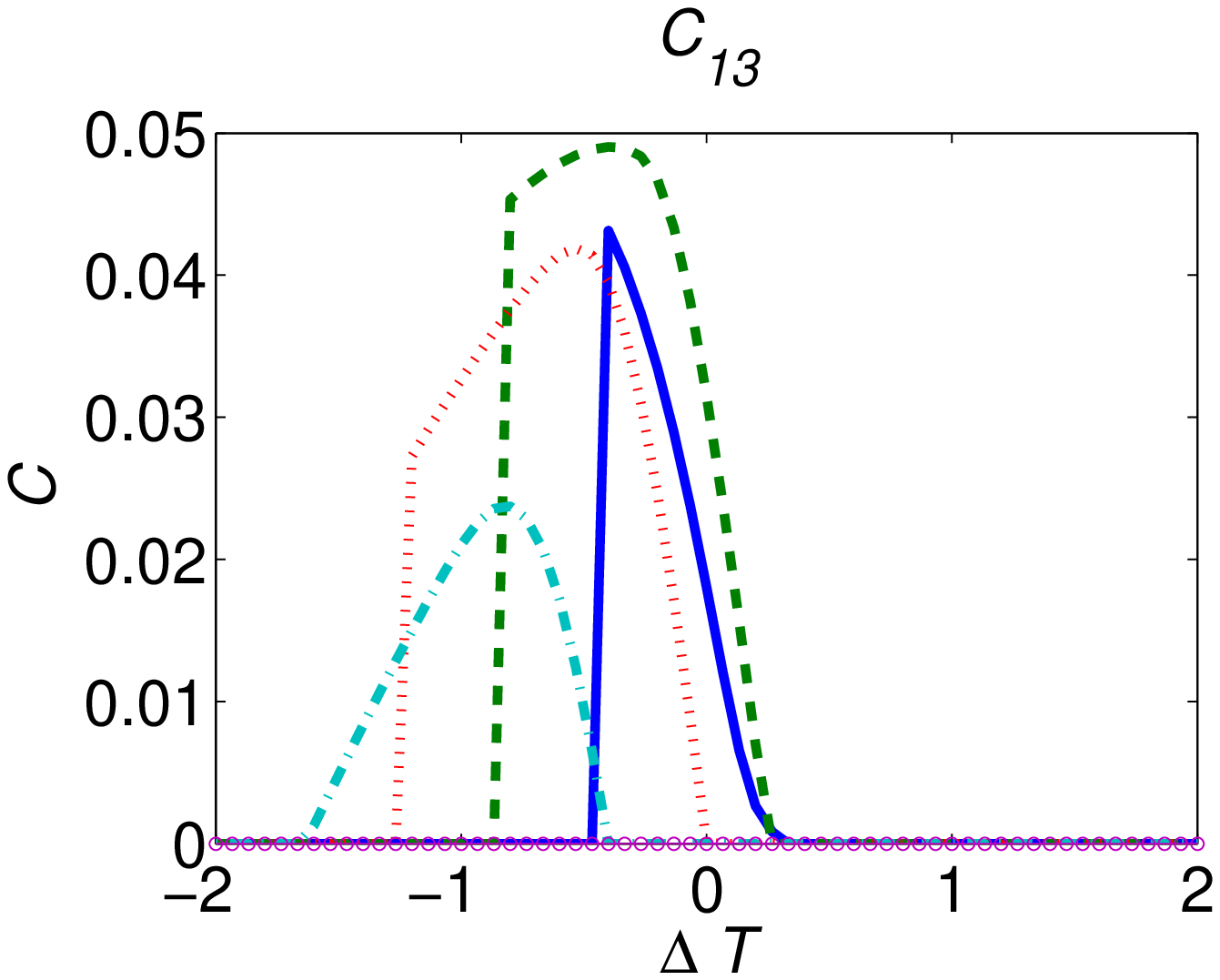}
\caption{(Color online) Left column, the steady state entanglement
as functions of the mean bath temperature $T_M{=}(T_L{+}T_R)/2$ and
the temperature difference $\Delta T{=}T_L{-}T_R$ in non-symmetrical
case. Right column shows the corresponding concurrences change with
$\Delta T$ in the case of $T_M{=}0.2$ (blue-solid), $T_M{=}0.4$
(green-dashed), $T_M{=}0.6$ (red-dotted), $T_M{=}0.8$
(cyan-dash-dot), and $T_M{=}1$ (pink-circle). Here
$\varepsilon{=}3$, $J_1{=}0.5$ and
$J_2{=}2.5$.}\label{FIG:Cw3J05d25TmdT}
\end{figure}

\begin{figure}
\includegraphics*[width=0.45\columnwidth,height=0.25\columnwidth]{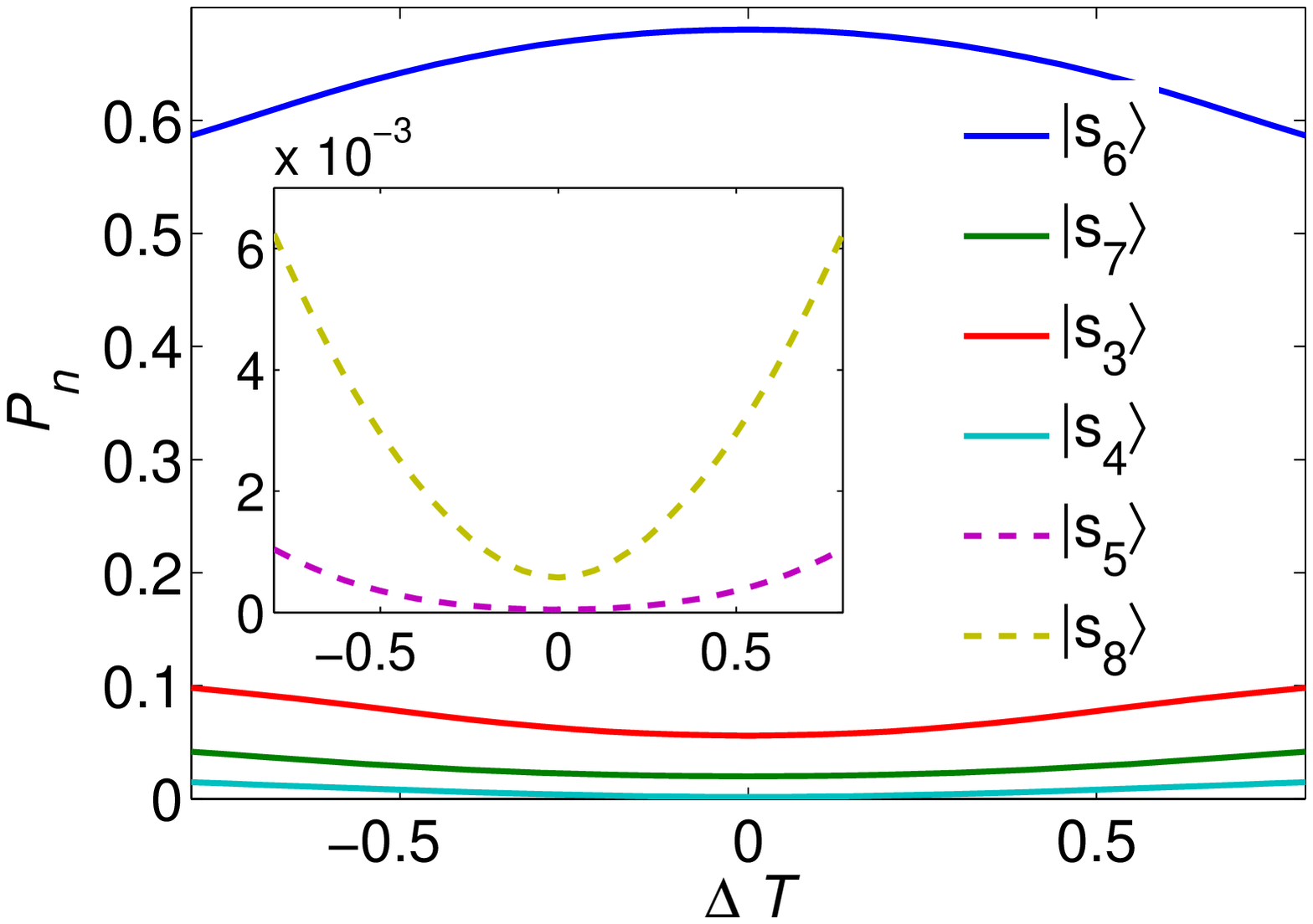}
\includegraphics*[width=0.45\columnwidth,height=0.25\columnwidth]{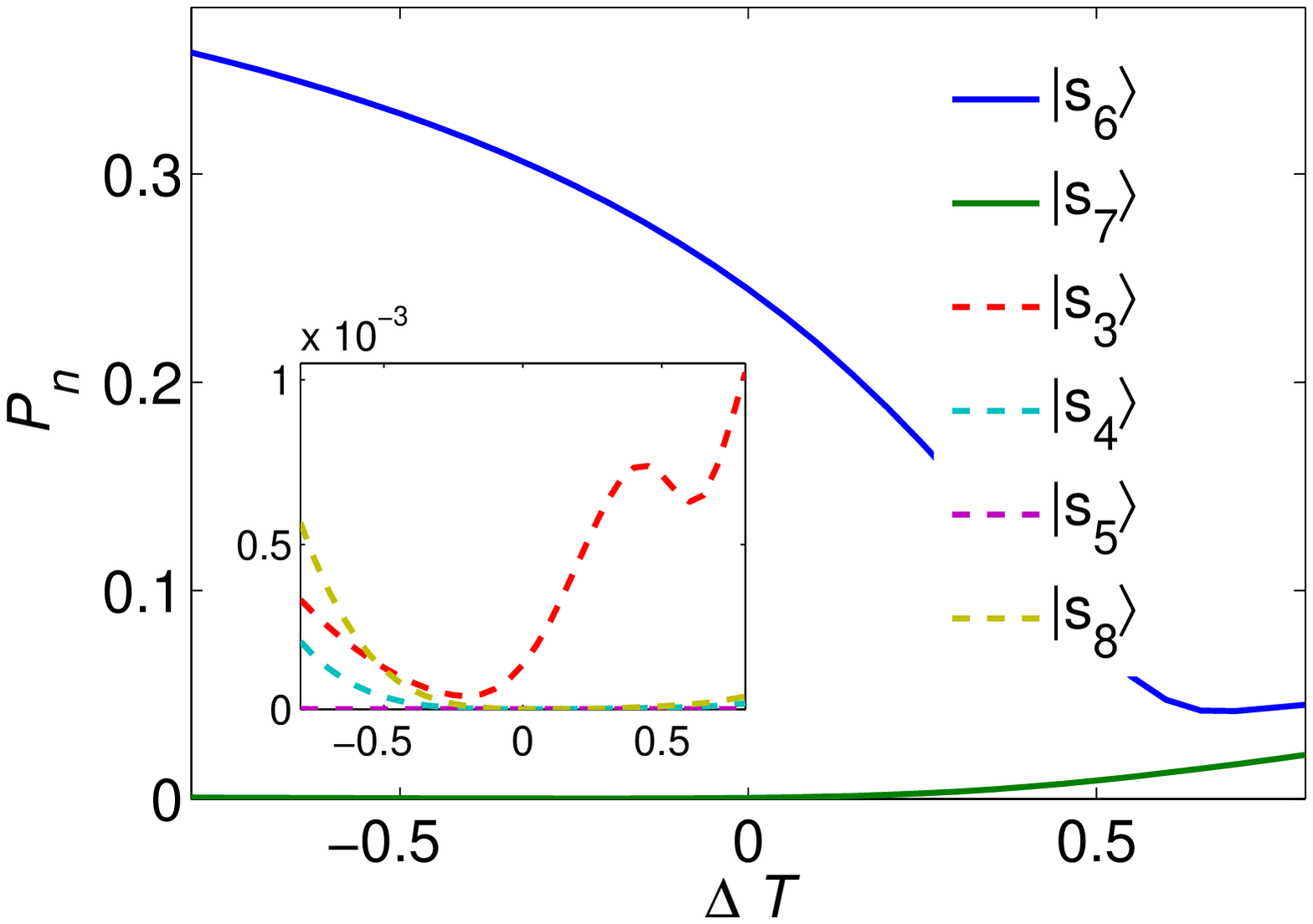}
\includegraphics*[width=0.8\columnwidth]{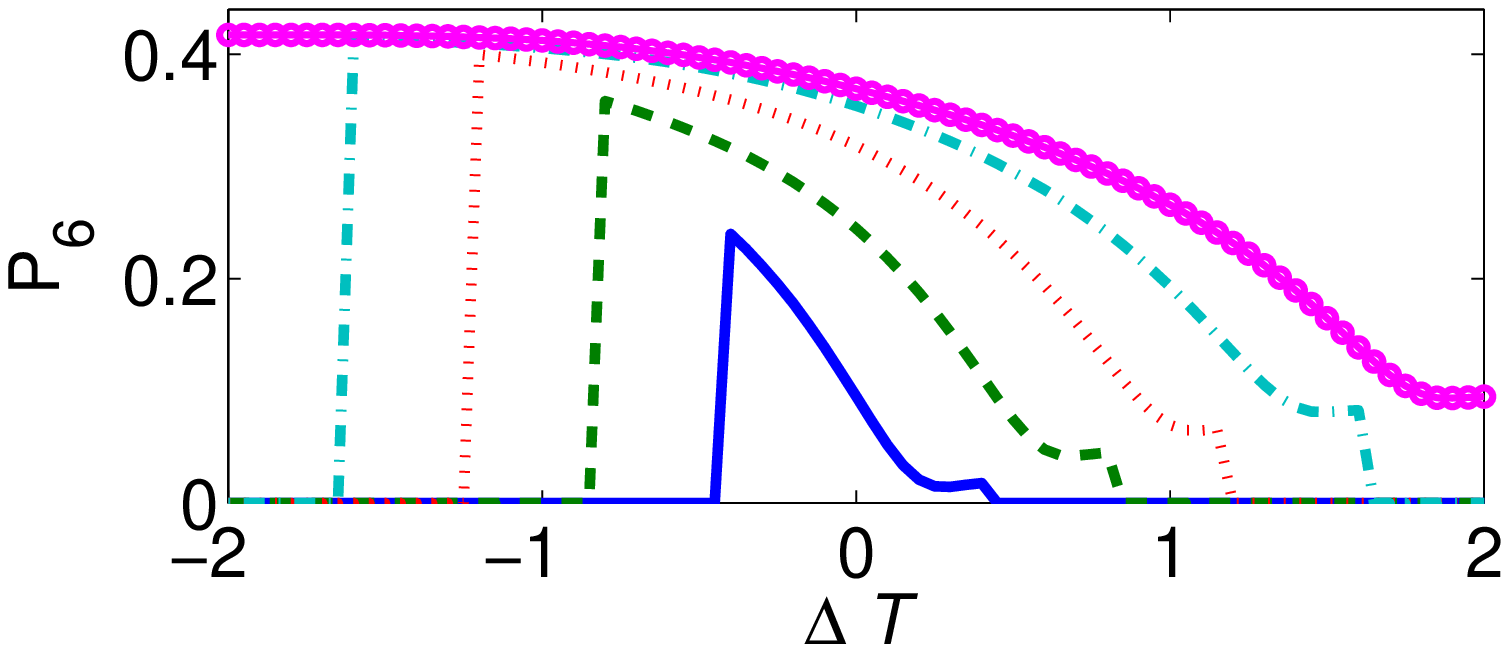}
\caption{(Color online) The probability distribution for eigenstates
$\ket{s_i},i{=}3,4,\cdots,8$ in the steady state as a function of
the temperature difference $\Delta T$. In top-left figure we set
$\varepsilon{=}1$ and $J_1{=}J_2{=}1$, while in top-right figure we
choose $\varepsilon{=}3$, $J_1{=}0.5$ and $J_2{=}2.5$. In both two
figures, the mean bath temperature $T_M$ is fixed with $T_M{=}0.4$.
In bottom figure, we plot the probability distribution for
$\ket{s_6}$ with different mean bath temperature. The mean bath
temperatures correspond to the right column of
Fig.\ref{FIG:Cw3J05d25TmdT}. }\label{FIG:distribution}
\end{figure}

\begin{figure}
\includegraphics*[bb=110 290 475 578, width=0.85\columnwidth]{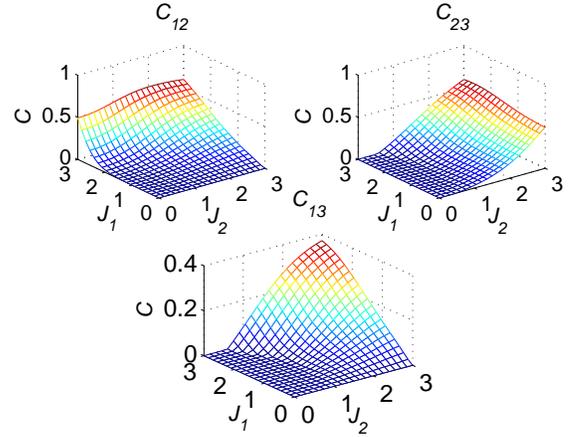}
\caption{Steady state entanglement as functions of the coupling
strength $J_1$ and $J_2$ in the case $\varepsilon{=}3$, $T_M{=}0.5$,
$\Delta T{=}-0.4$.}\label{FIG:Cw3T0307J1J2}
\end{figure}

Choosing  the concurrence \cite{concurrence} as the measure of
entanglement, we first study the steady state concurrence in the
case of  symmetric qubit-qubit couplings ($J_1{=}J_2$). In
Figs.\ref{FIG:Cw1J1} and \ref{FIG:Cw3J1}, we plot the steady state
concurrence $C_{12}$ (entanglement between spins 1 and 2), and
$C_{13}$ (entanglement between spins 1 and 3) at different
temperatures $T_L$ and $T_R$. In Fig.\ref{FIG:Cw1J1}, we set
$\varepsilon{=}1$ and in Fig.\ref{FIG:Cw3J1}, $\varepsilon{=}3$. The
coupling strength are chosen  to be $J_1{=}J_2{=}1$. Some features
can be observed from the figures. First, although the value of
$C_{12}$ and $C_{13}$ are quite different, the tendencies  of  them
affected by $T_L$ and $T_R$ are similar. The other interesting
feature  is that, in both Fig.\ref{FIG:Cw1J1} and
Fig.\ref{FIG:Cw3J1}, $C_{12}$ and $C_{13}$ are symmetric about
$T_L{=}T_R$, and the peak(maximum)  appears when $T_L{=}T_R$. This
means that, $C_{12}{=}C_{23}$ holds in this case. In other words,
the entanglements between the nearest neighbors are equal  although
the two temperatures are different. Moreover the larger the
temperature difference, the smaller the entanglement.

In nonsymmetric case ($J_1{\neq} J_2$), the results are quite
different. We first plot the concurrences between any two spins as
functions of the mean bath temperature $T_M{=}\frac12(T_L{+}T_R)$
and the temperature difference $\Delta T{=}T_L{-}T_R$ in
Fig.\ref{FIG:Cw3J05d25TmdT}. The parameters in the figure are chosen
$\varepsilon{=}3$, $J_1{=}0.5$ and $J_2{=}2.5$. It can be easily
seen from the figure that the peak does not appear at $\Delta T{=}0$
(For more clearly, see the right column of
Fig.\ref{FIG:Cw3J05d25TmdT}), i.e., for a fixed value of mean bath
temperature, the peak appears  at the points where the temperature
difference is not zero, suggesting that temperature difference
benefits the entanglement in nonsymmetric case. If one wants to get
a larger entanglement when the qubit-qubit couplings are not equal,
a specific  temperature difference is necessary. The other
interesting phenomenon in this case is that when the difference
between the coupling constant $J_1$ and $J_2$ is large enough (for
example, in our figure $J_1{=}0.5$, $J_2{=}2.5$), the difference
between $C_{12}$ and $C_{13}$ becomes negligible. In other words,
$C_{13}$, the concurrence between the next-to-nearest neighbor
qubits, tends to the concurrence between the nearest neighbors with
weak coupling (in our condition, $C_{12}$), and they all are smaller
than that in the case  with stronger couplings (in our condition,
$C_{23}$).


To understand these features, we calculate the probability
distribution
 for the eigenstates of the Hamiltonian $H$ in the steady
state, this distribution is  a function of temperature difference
$\Delta T$ as shown in Fig.\ref{FIG:distribution}. Although the
entanglement does not satisfy the role of
superposition\cite{entsuperposition}, the probability distribution
are helpful to understand the features we found in the entanglement.
 In the top-left figure of Fig.\ref{FIG:distribution}, we plot
this distribution for the symmetric case, i.e., $J_1{=}J_2=1$. The
other parameters are $\varepsilon{=}1$, $T_M{=}0.4$. From the
figure, we find that the probability distributions  for all the
eigenstates are symmetrical about $\Delta T{=}0$, and $\ket{s_6},
\ket{s_3}$ dominate the distribution. The distribution for
$\ket{s_1}$ and $\ket{s_2}$ are not shown in this figure, because
these two states are separable. On the other hand, for symmetric
couplings we have $\tan\theta{=}1$, then the eigenstates can be
divided  into two parts, i.e., symmetric eigenstates $\ket{s_3},
\ket{s_5}, \ket{s_6}, \ket{s_8}$, and antisymmetric eigenstates
$\ket{s_4}$ and $\ket{s_7}$. By symmetric we mean the state remains
unchanged when one exchanges  the particles 1 and 3, hence for
symmetric state
 the  entanglement between particles (1,2) and (2,3) equals.
 For the antisymmetric states, it is easy to check that
these is no entanglement between particle (1,2) or (2,3). These
observations  for the eigenstates together with their distribution
in the steady, we can conclude that the entanglement between the
nearest neighbors are equal although the two temperatures for the
two baths are different, this analysis confirms the numerical
simulation presented in the figure.

The top-right figure in Fig.\ref{FIG:distribution} shows the
probability distribution for the nonsymmetric case. We have set
$\varepsilon{=}3$, $J_1{=}0.5$ and $J_2{=}2.5$. From this figure, we
find that entanglement in the steady states is mainly determined by
the state $\ket{s_6}$. We plot the probability distribution of
$\ket{s_6}$ with different mean bath temperature in the bottom of
Fig.\ref{FIG:distribution}. Now we analyze the entanglement
properties for this state. By tracing out one particle from the
state $\ket{s_6}$, we can easily obtain the concurrence for the
remaining two spins as $C_{12}{=}\cos\theta$,
$C_{13}{=}\cos\theta\sin\theta$ and $C_{23}{=}\sin\theta$. In
nonsymmetric case, for example, $J_1{=}0.5,J_2{=}2.5$, we have
$\tan\theta{=}5$, which results in $\sin\theta{\simeq} 1$. This is
the reason why  $C_{12}{\simeq} C_{13}$ in
Fig.\ref{FIG:Cw3J05d25TmdT} and they are much smaller than $C_{23}$.
Moreover, observing  the distribution probability for $\ket{s_6}$
(blue line in the middle figure of Fig.\ref{FIG:distribution} and
the bottom figure), we can find  that the peak of entanglement does
not appear at $\Delta T{=}0$ in nonsymmetric case, namely the
temperature difference favors the steady state entanglement.

Finally we study the effect of two coupling constant on the
concurrence for fixed bath temperature. In
Fig.\ref{FIG:Cw3T0307J1J2}, we show the concurrences as functions of
$J_1$ and $J_2$ with $\varepsilon{=}3$, $T_M{=}0.5$ and $\Delta
T{=}-0.4$. Both $J_1$ and $J_2$ enhance the entanglement. This
enhancement is more strikingly in the case of $J_1$ for $C_{12}$
while $J_2$ for $C_{23}$. Due to the temperature difference,
$C_{13}$ is not symmetric about $J_1{=}J_2$.

In summary, we have studied the steady state entanglement in a
three-qubit $XX$ model. The qubits are coupled to two  independent
bosonic baths at different temperatures.  With the help of the
effective Hamiltonian approach, we have calculated the steady state
entanglement and discussed its dependence on temperatures and
couplings. Two types of the coupling, i.e., symmetric  and
nonsymmetric one are considered. When the coupling is symmetric, we
find that the entanglement  between the nearest neighbors, i.e.,
$C_{12}$ and $C_{23}$, are equal though the temperatures of the two
baths are different. The maximum of entanglement  is found when the
temperatures of the two baths are equal. For the nonsymmetric case,
however, the maximal entanglement arrives when the two baths at
different temperatures. By analyzing the distribution for each
eigenstate, we qualitatively explain these interesting phenomena.
The dependence of the entanglement on the coupling constants are
also presented and discussed.

We thank Dr H. T. Cui for discussion. This work is supported by NSF
of China under grant Nos. 10775023 and 10935010.

\end{document}